\documentclass[aps,prb]{revtex4}
\usepackage{amsmath,amssymb}
\usepackage{graphicx}
\usepackage{bm}
\usepackage{color}
\newcommand{\mbb}{\mathbb}
\newcommand{\mc}{\mathcal}

\newcommand{\tet}{\texttt}
\newcommand{\pr}{\partial}

\begin{document}
\title{Nonlinear transport of ballistic Dirac electrons tunneling through a tunable potential barrier in graphene}

\author{Farhana Anwar$^{1,2}\footnote{fanwar@unm.edu}$, Andrii Iurov$^{3,1}\footnote{aiurov@mec.cuny.edu, theorist.physics@gmail.com}$, 
Danhong Huang$^{4}$, Godfrey Gumbs$^{5,6}$ and Ashwani Sharma$^{1,2,4}$}
\affiliation{$^{1}$Center for High Technology Materials, University of New Mexico,
1313 Goddard SE, Albuquerque, NM 87106, USA\\
$^{2}$Department of Electrical and Computer Engineering, University of New Mexico, Albuquerque, NM 87106, USA\\
$^{3}$Department of Physics and Computer Science, Medgar Evers College of City University of New York, Brooklyn, NY 11225, USA\\
$^{4}$Air Force Research Laboratory, Space Vehicles Directorate,
Kirtland Air Force Base, NM 87117, USA\\
$^{5}$Department of Physics and Astronomy, Hunter College of the City
University of New York, 695 Park Avenue, New York, NY 10065, USA\\
$^{6}$Donostia International Physics Center (DIPC),
P de Manuel Lardizabal, 4, 20018 San Sebastian, Basque Country, Spain}

\date{\today}

\begin{abstract}
Dirac-electronic tunneling and nonlinear transport properties with both finite and zero
energy bandgap are investigated for graphene with a tilted potential barrier under a bias. 
For validation, results from a finite-difference based numerical approach, which is developed for calculating transmission and reflection coefficients with a dynamically-tunable 
(time-dependent bias field) barrier-potential profile,  
are compared with those of both an analytical model for a static square-potential barrier and a perturbation theory using Wentzel-Kramers-Brillouin (WKB) approximation. For a biased barrier, both 
transmission coefficient and tunneling resistance are computed and analyzed, 
indicating a full control of the peak in tunneling resistance 
by bias field for a tilted barrier, gate voltage 
for barrier height, and energy for incoming electrons.
Moreover, a finite energy gap in graphene is found  to suppress head-on transmission
as well as skew transmission with a large transverse momentum. For a gapless graphene, 
on the other hand, filtering of Dirac electrons outside of normal incidence is found 
and can be used for designing electronic lenses. 
All these predicted attractive transport properties are expected extremely useful for the development of novel electronic and optical graphene-based devices. 
\end{abstract}

\maketitle

\section{Introduction}
\label{s1}

Graphene, a one atom-thick allotrope of carbon, being a conductor with exceptionally large mobility at a large 
range of ambient temperatures, makes a strong case for a number of ballistic transport nanodevices. It has unique 
electronic properties due to its linear energy dispersion with zero bandgap, as well as a spinor two-component wave function. 
These unique characteristics give rise to some highly unusual electronic and transport properties.\,\cite{kats,neto,sarma2011electronic,novoselov2005two}
\medskip 

These peculiar properties result in a fact that a potential barrier becomes transparent to electrons arriving at normal 
incidence regardless of its height or width. This effect, known as Klein tunneling,\,\cite{kats} restricts the switching-off
capability (complete pinch-off of electric current) for logical applications, and makes graphene difficult to achieve logical functionalities without 
use of chemical modification or patterning.\,\cite{wang2019graphene,low2009electronic, jang2013graphene,wilmart2014klein}
\medskip

On the other hand, such a situation also offers a unique possibility to fabricate various ballistic devices and circuits in which 
electrons experience focusing by one or several potential barriers. Practically, a zero-bandgap two-dimensional material acquires an important
advantage over metals in its capability to tune the conductivity by means of either chemical doping or a gate voltage
with a desired geometrical pattern.\,\cite{novoselov2006unconventional,zhang2005experimental,ohta2006controlling}
\medskip 

Interestingly, the induced planar barrier structure within a grraphene sheet can be realized by applying a gate voltage, either a static or a transient one.
This is quite different from the design of a high-electron-mobility transistor.
For example, by using different inhomogeneous profiles of static bias voltage, various structures , such as bipolar ($p-n$, $n-p$, $p-n-p$, $n-p-n$, etc.) as 
well as unipolar junctions ($n-n'$, $p-p'$) can be facilitated\,\cite{cheianov2006selective,shytov2008klein,sonin2009effect,chen2016electron,cayssol2009contact,allain2011klein,phong2016fermionic}
to achieve desired voltage dependence of electrical conductance. In spite of the considered junction being abrupt or graded, its angular selectivity for carrier transport
makes it a unique one in comparison with conventional semiconductor junctions, e.g., metal–oxide–silicon field-effect transistors.
\medskip
 
Other significant roles played by our proposed tunable junctions include Veselago
lens,\,\cite{cheianov2007focusing,dip} Fabry-Perot interferomete,r\,\cite{shytov2008klein} subthermal 
switches,\,\cite{wang2019graphene} Andreev reflections,\,\cite{beenakker2006specular} by exploiting optics-like behavior of
ballistic Dirac electrons. Therefore, in order to design future-generation graphene-based electronics, it is crucial to gain a full
understanding of mechanism for ballistic transports across various types of potential barriers in graphene. 
Negative refractive index with a single ballistic graphene junction, which is associated with electron-hole switching, 
has already been observed experimentally,\,\cite{chen2016electron,wang2019graphene} 
and it strongly affects the operation of an electric switch.\,\cite{sajjad2011high,elahi2019impact,wang2019graphene}
\medskip 

Various theoretical methods have been adopted aiming to obtain electron transport in graphene, including 
transfer matrix,\,\cite{tworzydlo2008finite,hernandez2012finite,xu2010resonant} non-equilibrium
Green's function,\,\cite{low2009conductance,low2009electronic} tight-binding model,\,\cite{logemann2015modeling,
ghobadi2013device} as well as semi-classical Wentzel-Kramers-Brillouin (WKB) approximation.\,\cite{sonin2009effect,
allain2011klein,logemann2015modeling} However, there are still few of studies on electronic transport properties using 
finite-difference method\,\cite{huang1999effects} for numerical calculation with an arbitrary potential profile. 
A crucial advantage of this numerical method is a possibility to take into account of random local disorder potentials within barrier materials 
of roughness at two barrier edges.
A number of fabricated optical devices face such a situation, which detriments the device performance\,\cite{elahi2019impact,wilmart2014klein} while trying to accomplish 
ballistic $p-n$ junction characteristics experimentally. 
Alternatively, some smooth $p-n$ and $n-p-n$ junctions in graphene were realized and analyzed
theoretically in Refs.\,[\onlinecite{jang2013graphene,cheianov2006selective, shytov2008klein, oura, chen2016electron, wilmart2014klein, ourt}].
\medskip 

The remaining part of this paper is organized as follows. In Sec.\,\ref{s2}, we introduce finite-difference method for calculating transmission coefficient of 
Dirac electrons in graphene in the presence of a biased potential barrier, along with numerical results of transmission coefficient as functions of incident angles and electron energy, 
as well as tunneling resistance as a function incident electron energy, with various values of bias field. 
We present in Sec.\,\ref{s3} analytical results within the WKB approximation for both large and small bias-field limits, accompanied by numerical results for  
transmission coefficient as functions of both incident angles and electron energy.
Finally, our concluding remarks are presented in Sec.\,\ref{s4}.

\section{Finite-difference method for tunneling of Dirac electrons}
\label{s2}

In this Section, we lay out the formalism, and present and discuss our numerical results based on a finite-difference method. 
The main advantage of this method is its capability to obtain exact 
electron wave functions for arbitrary potential profiles.\,\cite{huang1999effects} 
\medskip 

We will consider both cases with a zero or finite energy gap for graphene. 
Technically, an energy gap ($\sim 200\,$meV) could be introduced by placing a graphene 
sheet on top of either insulating silicon-based\,\cite{p7} or hexagonal boron-nitride substrate.\,\cite{p777} 
It could also be realized by patterned hydrogen adsorption\,\cite{p77} or imposing a circularly-polarized 
off-resonance laser field.\cite{kibis1, kibis2} This gap opening leads to substantial modifications of 
electronic, transport and collective properties of graphene, e.g., plasmon dispersions.\,\cite{hb, pl16, pavlo1}

\subsection{Electronic States of Gapped Graphene}
\label{s2-1}

For gapped graphene, there exists a finite energy bandgap $E_G=2\Delta_G$ between the valence and conduction 
bands with energy dispersion $\varepsilon_\gamma(k) = \gamma \sqrt{(\hbar v_F k)^2 + \Delta_G^2}$, where 
$\gamma = \pm 1$ correspond to electron and hole state, respectively. 
The Hamiltonian matrix associated with this dispersion possesses an additional $\hat{\Sigma}_z$ term on top of the 
Dirac Hamiltonian for gapless graphene,\,\cite{pavlo1, oura} yielding

\begin{equation}
\label{generalHam}
\hat{\mc{H}}_g(\mbox{\boldmath$r$})=-iv_F\,\hat{\mbox{\boldmath$\Sigma$}}_{x,y}\cdot \mbox{\boldmath$\nabla$}_{\bf r}
+V_B(x)\,\hat{\Sigma}_{0} + \Delta_G\,\hat{\Sigma}_z\ , 
\end{equation}
where $\mbox{\boldmath$r$}=(x,y)$,
$\hat{\Sigma}_{x,y,z}$ are two-dimensional Pauli matrices, $\hat{\Sigma}_{0}$ is a $(2 \times 2)$ unit matrix,
and $V_B(x)$ is a spatially-nonuniform barrier potential.
\medskip 

In general, the scattering-state solution for the Hamiltonian in Eq.\,\eqref{generalHam} has a
two-component (spinor) type of wave function $\Psi_{\gamma}(\mbox{\boldmath$r$})= \tet{exp}(ik_yy)\,\Phi_{\gamma}(x)=\, \tet{exp}(ik_yy)\,\left[\phi_A^{\gamma}(x),\,\phi_B^{\gamma}(x)\right]^T$, where $\gamma=\text{sign}[\varepsilon_0(k)-V_B(x)]=\pm 1$ represents the electron-hole index and 
$\varepsilon_0(k)$ is the given energy of an incident electron. 
\medskip 

For the case with a constant barrier potential $V_0$, however, the Hamiltonian in Eq.\,\eqref{generalHam} can be greatly simplified as

\begin{equation}
\label{gap01}
\hat{\mc{H}}^{(0)}_g(k \, \vert \, \theta_{\bf k}) = \left[
\begin{array}{cc}
V_0+\Delta_G & \hbar v_Fk_-\\
\\
\hbar v_Fk_+ & V_0-\Delta_G
\end{array}
\right]\ , 
\end{equation}
where $\mbox{\boldmath$k$}=(k_x^{(0)},k_y)$ and $k_{\pm} = k_x^{(0)} \pm i k_y$. In this case, the scattering-state wave function related to the Hamiltonian in Eq.\,\eqref{gap01} gains the explicit form\,\cite{oura,ourjap}

\begin{equation}
\Psi_{\gamma}^{(0)}(\mbox{\boldmath$r$}) =\frac{1}{\sqrt{2\gamma\,\delta\varepsilon_0(k)}} \left[
\begin{array}{c}
\sqrt{\vert \delta\varepsilon_0(k)+\Delta_G\vert}\\
\\
\gamma\sqrt{\vert\delta\varepsilon_0(k)-\Delta_G\vert}\,\tet{e}^{i\theta_{\bf k}}
\end{array}
\right]\tet{exp}(ik^{(0)}_xx+ik_yy)\ ,
\label{wfunc}
\end{equation}
where $\theta_{\bf k}=\tan^{-1}(k_y/k_x^{(0)})$, $\delta\varepsilon_0(k)\equiv\varepsilon_0(k)-V_0\geq\Delta_G$ for $\gamma=+1$, while 
$\delta\varepsilon_0(k)\leq-\Delta_G$ for $\gamma=-1$. Here, two components of the wave function in Eq.\,\eqref{wfunc}
are not the same but they are still interchangeable for electrons and holes with $\gamma = \pm 1$. 

\subsection{Finite-Difference Method for Tunneling Dirac Electrons}
\label{s2-2}

For the Hamiltonian in Eq.\,\eqref{generalHam}, a pair of scattering-state equations within the barrier region are obtained as  

\begin{eqnarray}
\nonumber 
&& \frac{d\phi_B(x)}{dx}+k_y\,\phi_B(x)=\frac{i}{\hbar v_F}\,\left[\varepsilon_0(k)-V_B(x)+V_D\,\delta(x-x_D)-\Delta_G\right]\phi_A(x)\ , \\
\label{fd1}
&& \frac{d\phi_A(x)}{dx}-k_y\,\phi_A(x)=\frac{i}{\hbar v_F}\,\left[\varepsilon_0(k)-V_B(x)+V_D\,\delta(x-x_D)+\Delta_G\right]\phi_B(x)\ .  
\end{eqnarray} 
Here, we consider a titled potential barrier under an applied electric 
field ${\cal E}_{dc}$, which gives rise to $V_B(x)=V_0-e{\cal E}_{dc}\,x$ in the barrier region,
where $V_0$ and ${\cal E}_{dc}$ can be either positive or negative. Additionally, $k_y$ of electrons remains conserved during a tunneling process along the $x$ direction.
Moreover, a single disorder at $0<x=x_D<W_B$ is assumed within the barrier region with 
a constant trap-potential amplitude $-V_D$.
\medskip

Mathematically, we can divide the electron wave function corresponding to three separated regions. To the left of the potential barrier $x<0$, we acquire the wave function

\begin{equation}
\label{wfleft}
\Phi_<(x)
=s(\varepsilon_0)
\left[
\begin{array}{c}
1\\
\\
\tet{e}^{i\theta_{\bf k}} 
\end{array}
\right]
\tet{exp}(ik^{(0)}_xx)+
r(\varepsilon_0)
\left[  
\begin{array}{c}
1\\
\\
-\tet{e}^{i\theta_{\bf k}} 
\end{array}
\right]
\tet{exp}(-ik^{(0)}_xx)\ ,
\end{equation} 
where $s(\varepsilon_0)$ and $r(\varepsilon_0)$ represent incoming and reflected wave-function amplitudes. To the right of the potential barrier $x>W_B$, on the other hand, the wave function
is found to be

\begin{equation}
\label{wfright}
\Phi_>(x)
=t(\varepsilon'_0)
\left[
\begin{array}{c}
1\\
\\
\tet{e}^{i\theta_{{\bf k}'}} 
\end{array}
\right]
\tet{exp}(ik'_xx)\ ,
\end{equation}
where $t(\varepsilon'_0)$ is the transmitted wave-function amplitude.
\medskip

Results in Eqs.\,\eqref{wfleft} and \eqref{wfright} can be applied to construct boundary conditions on both sides of a potential barrier. 
For the wave function within the barrier region, a finite-difference method can be employed to seek for a numerical solution of Eq.\,\eqref{fd1}.
Following the procedure adopted in Ref.\,[\onlinecite{huang1999effects}] for a two-dimensional electron gas, we discrete the whole barrier
region $0\leq x\leq W_B$ into $N_B$ (odd integer) equally spaced slabs, and each slab has the same width $\Delta_0=W_B/N_B$. Therefore, two coupled
differential equations in Eq.\,\eqref{fd1} can be solved simultaneously through a backward-iteration procedure in combination with two continuity boundary conditions at $x=W_B$ and $x=0$. Especially, for $\Delta_G=0$ we find the following backward iterative relation for $ 1\leq j \leq N_B+1$ and $x_j=(j-1)\Delta_0$

\begin{equation}
\label{fd2} 
\left\{
\begin{array}{c}
\phi_{A}(x_{j-1})\\
\\
\phi_{B}(x_{j-1})
\end{array}
\right\}= 
\left\{
\begin{array}{c}
\phi_{A}(x_{j})\\
\\
\phi_{B}(x_{j})
\end{array}
\right\}
-k_y\Delta_0
\left\{
\begin{array}{c}
\phi_{A}(x_{j})\\
\\
-\phi_{B}(x_{j})
\end{array}
\right\}
+\frac{i\Delta_0}{\hbar v_F}\,
\big[
\varepsilon_0(k)-V_0+e{\cal E}_{dc}x_j+V_D\delta(x_j-x_D)
\big]\left\{
\begin{array}{c}
\phi_{B}(x_{j})\\
\\
\phi_{A}(x_{j})
\end{array}
\right\}\ .
\end{equation}
\medskip

By using Eq.\,\eqref{wfright}, the first continuity boundary condition at 
$x_{N_B+1}=W_B=N_B\Delta_0$ leads to

\begin{equation}
\label{NBp1}
\left\{
\begin{array}{c}
\phi_{A}(x_{N_B+1})\\
\\
\phi_{B}(x_{N_B+1})
\end{array}
\right\}
= t(\varepsilon'_0(k'))\, 
\left\{
\begin{array}{c}
1\\
\\
\tet{e}^{i\theta_{{\bf k}'}} 
\end{array}
\right\}
\tet{exp}(ik'_xN_B\Delta_0)
\,\tet{exp}\left[k_y\Delta_0\sum\limits_{j=2}^{N_B+1}\,\Theta(-\kappa(x_j))/\sqrt{\vert\kappa(x_j)\vert}\right]_{x_j\neq x_D}\ ,
\end{equation}
where $\kappa(x_j)=\left[(1/\hbar v_F)\left(\varepsilon_0(k)-V_0+e{\cal E}_{dc}x_j\right)\right]^2-k_y^2$, $\varepsilon'_0(k')=\hbar v_F\sqrt{k_x^{'2}+k_y^2}=\varepsilon_0(k)+e{\cal E}_{dc}W_B$, 
and $\Theta(x)$ is a step function, i.e., $\Theta(\xi)=1$ for $\xi>0$ while zero for others. 
Physically, the last exponential term in Eq.\,\eqref{NBp1} does not affect the transmission coefficient if $\kappa(x_j)>0$, corresponding to a semi-classical regime. However, this term can significantly reduce the transmission coefficient, but not the reflection coefficient, if $\kappa(x_j)<0$, connecting to a quantum-tunneling regime. 
The backward iteration in Eq.\,\eqref{fd2} can be performed all the way down to $x_1=0$. 
\medskip

In a similar way, using Eq.\,\eqref{wfleft} and another continuity boundary condition at $x_1=0$, we find 

\begin{equation}
\left\{
\begin{array}{c}
\vert s(\varepsilon_0)\vert^2\\
\\
\vert r(\varepsilon_0)\vert^2
\end{array}
\right\}= 
\frac{1}{4}
\left\{
\begin{array}{c}
\vert a\vert^2+\vert b\vert^2+2\tet{Re}(ab^*\tet{e}^{i\theta_{{\bf k}}})\\
\\
\vert a\vert^2+\vert b\vert^2-2\tet{Re}(ab^*\tet{e}^{i\theta_{{\bf k}}})
\end{array}
\right\}\ ,
\end{equation}
where we have defined the notations

\begin{equation}
\left\{
\begin{array}{c}
a\\
\\
b
\end{array} 
\right\}
\equiv
\left\{
\begin{array}{c}
\phi_{A}(x_1)\\
\\
\phi_{B}(x_1)
\end{array} 
\right\}=
s(\varepsilon_0) 
\left\{
\begin{array}{c}
1\\
\\
\tet{e}^{i\theta_{{\bf k}}}
\end{array}
\right\} 
+r(\varepsilon_0)
\left\{
\begin{array}{c}
1\\
\\
-\tet{e}^{i\theta_{{\bf k}}} 
\end{array}
\right\}\ .
\label{coff}
\end{equation} 
\medskip

The transmission coefficient $T(k,\theta_{\bf k} \, \vert \,{\cal E}_{dc})$, which is defined as the ratio of the transmitted to the incident 
probability current densities,\,\cite{neto, kats} is 
given by 

\begin{equation}
\label{trans}
T(k,\theta_{\bf k} \, \vert \,{\cal E}_{dc})=\frac{\vert t(\varepsilon'_0)\vert^2}
{\vert s(\varepsilon_0)\vert^2}\ ,
\end{equation}
since electrons on both sides of the potential barrier have the same group velocity $v_F$.
Numerically, it is easy to set $t(\varepsilon'_0)\equiv 1$, then to find $s(\varepsilon_0)$ through Eq.\,\eqref{coff} after having performed all the backward iterations, and finally obtain the ratio in Eq.\,\eqref{trans}.
Using the transmission coefficient in Eq.\,\eqref{trans}, we are able to compute the tunneling electric current 
$J_0$ per length, yielding

\begin{equation}
J_0=\frac{4e}{{\cal A}}\sum\limits_{\bf k}\,T(k,\theta_{\bf k} \, \vert \,{\cal E}_{dc})\,
v_F\cos\theta_{\bf k}\left[f_0(\varepsilon_0(k))-f_0(\varepsilon_0(k)+e{\cal E}_{dc}W_B)\right]\ ,
\end{equation}
where ${\cal A}$ is the graphene sheet area, $f_0(x)=\left\{1+\exp\left[(x-u_0)/k_BT\right]\right\}^{-1}$ is the Fermi function for thermal-equilibrium electrons at temperature $T$, and $u_0(T)$ is the chemical potential of electrons. For a weak electric field, we have $e{\cal E}_{dc}W_B\ll\varepsilon_0(k)$, which leads to

\begin{equation}
J_0\approx\frac{4e^2v_FU_0}{{\cal A}}\sum\limits_{\bf k}\,T(k,\theta_{\bf k} \, \vert \,{\cal E}_{dc})\,
\cos\theta_{\bf k}\left[-\frac{\partial f_0(\varepsilon_0(k))}{\partial\varepsilon_0}\right]\ ,
\end{equation}
were $U_0={\cal E}_{dc}W_B$ represents the voltage drop across the potential barrier.
If $T$ is low, i.e., $k_BT\ll E_F$ with $E_F$ as the zero-temperature $u_0$ or Fermi energy, we find

\begin{eqnarray}
\nonumber
J_0&\approx&\frac{4e^2v_FU_0}{{\cal A}}\sum\limits_{\bf k}\,T(k,\theta_{\bf k} \, \vert \,{\cal E}_{dc})\,
\cos\theta_{\bf k}\,\delta(\hbar v_Fk-E_F)
=\frac{U_0}{\pi}\left(\frac{2e^2}{h}\right)\int\limits_0^{\infty} dk\,k\,\delta(k-k_F)\int\limits_{-\pi/2}^{\pi/2} d\theta_{\bf k}\,T(k,\theta_{\bf k} \, \vert \,{\cal E}_{dc})\,
\cos\theta_{\bf k}\\
&=&{\cal E}_{dc}\,k_FW_B\left(\frac{2e^2}{h}\right)\left\{\frac{1}{\pi}\int\limits_{-\pi/2}^{\pi/2} d\theta_{\bf k}\,T(k_F,\theta_{\bf k} \, \vert \,{\cal E}_{dc})\,
\cos\theta_{\bf k}\right\}\ ,
\end{eqnarray}
where $k_F=\sqrt{\pi n_0}$ is the Fermi wave vector and $n_0$ is the areal electron density.
Finally, we obtain the nonlinear two-terminal sheet tunneling conductivity $\sigma(k_F,{\cal E}_{dc})$ (in units of $2e^2/h$), given by\,\cite{cond1} 

\begin{equation}
\sigma(k_F,{\cal E}_{dc})=\frac{J_0}{{\cal E}_{dc}}=\frac{k_FW_B}{\pi}\int\limits_{-\pi/2}^{\pi/2} d\theta_{\bf k}\,T(k_F,\theta_{\bf k} \, \vert \,{\cal E}_{dc})\,
\cos\theta_{\bf k}\ .
\label{cond}
\end{equation} 
Specifically, for normal incidence of electrons with $\theta_{\bf k}\equiv 0$, we simply get
$\sigma_0(k_F,{\cal E}_{dc})=(k_FW_B)\,T(k_F \,\vert\,{\cal E}_{dc})$.
\medskip

To simulate disorder effects on the tunneling of Dirac electrons, we introduce a normal distribution function and replace the transmission coefficient 
$T(k_F,\theta_{\bf k}\,\vert\,{\cal E}_{dc})\equiv T(k_F,\theta_{\bf k}\,\vert\,{\cal E}_{dc},x_D)$ 
in Eqs.\,\eqref{trans} and \eqref{cond} by its average $\overline{T}(k,\theta_{\bf k}\,\vert\,{\cal E}_{dc})$, yielding

\begin{equation}
\overline{T}(k,\theta_{\bf k}\,\vert\,{\cal E}_{dc})=\frac{1}{N_D}\int\limits_0^{W_B} dx_D\,
T(k_F,\theta_{\bf k}\,\vert\,{\cal E}_{dc},x_D)\,\rho(x_D\,\vert\,\sigma_0)
\approx\frac{\Delta_0}{N_D}\sum\limits_{s=2}^{N_B}\,T(k_F,\theta_{\bf k}\,\vert\,{\cal E}_{dc},x^*_s)\,\rho(x^*_s\,\vert\,\sigma_0) \ ,
\label{dist}
\end{equation}
where $x^*_s=(s-1)\Delta_0$, the introduced distribution function is assumed to be

\begin{equation}
\rho(x^*_s\,\vert\,\sigma_0)=\frac{1}{\sqrt{2\pi\sigma_0^2}}\,
\exp\left[-\frac{(x^*_s-W_B/2)^2}{2\sigma_0^2}\right]
\end{equation}
with the standard deviation $\sigma_0=\Delta_0$ and $W_B/2=[(N_B+1)/2]\Delta_0$. In addition, the normalization factor in Eq.\,\eqref{dist} is given by

\begin{equation}
N_D=\int\limits_0^{W_B} dx_D\,\rho(x_D\,\vert\,\sigma_0)\approx\Delta_0\sum\limits_{s=2}^{N_B}\,\rho(x^*_s\,\vert\,\sigma_0)\ .
\end{equation}
For convenience, in numerical calculations we further approximate the delta-function in Eq.\,\eqref{fd2} by

\begin{equation}
\delta(x_j-x_D)\equiv\delta(x_j-x^*_s)\approx\frac{\Gamma/\pi}{(x_j-x^*_s)^2+\Gamma^2}
\end{equation}
with a broadening parameter $\Gamma=\Delta_0$.

\subsection{Results for Dirac-Electron Tunneling in Graphene}
\label{s2-3}

\begin{figure}
\centering
\includegraphics[width=0.7\textwidth]{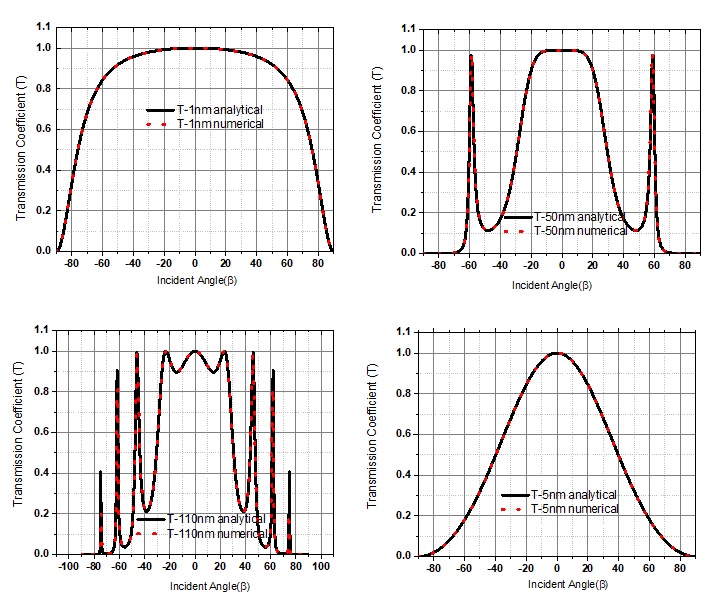}
\caption{(Color online) Comparison of calculated transmission coefficients $T(\varepsilon,\phi_{\bf k}\,\vert\,{\cal E}_{dc}=0)$ as a function of incident angle $\phi_{\bf k}$ 
based on either an analytical solution (black-solid curves) or finite-difference approach (red-dashed curves) for four different barrier thickness $W_B=1\,$nm (upper-left), $50\,$nm (upper-right), 
$110\,$nm (lower-left) and $5\,$nm (lower-right),
where $V_0=285\,$meV and $\varepsilon/V_0=1.25$ are chosen for calculations.}
\label{f2}
\end{figure} 

In the previous subsections\ \ref{s2-1} and \ref{s2-2}, 
we have established a finite-difference numerical scheme and applied it to study tunneling transport of carriers through a biased potential barrier in graphene. 
As a validation, we first compare our finite-difference results with those from an analytical solution\,\cite{neto} for a square potential barrier $V_B(x)=V_0$. 
Figure\ \ref{f2} displays a comparison for calculated transmission coefficients $T(\varepsilon,\phi_{\bf k}\,\vert\,{\cal E}_{dc}=0)$ as a function of incident angle $\phi_{\bf k}$ 
using either an analytical solution\,\cite{neto} (black solid curves) or our finite-difference method presented in subsection\ \ref{s2-2} (red dashed curves).
The results in this figure clearly indicate that our finite-difference method in subsection\ \ref{s2-2} is valid and can be applied to arbitrary potential profiles $V_B(x)$ 
including a biased potential barrier.
\medskip

\begin{figure}
\centering
\includegraphics[width=0.9\textwidth]{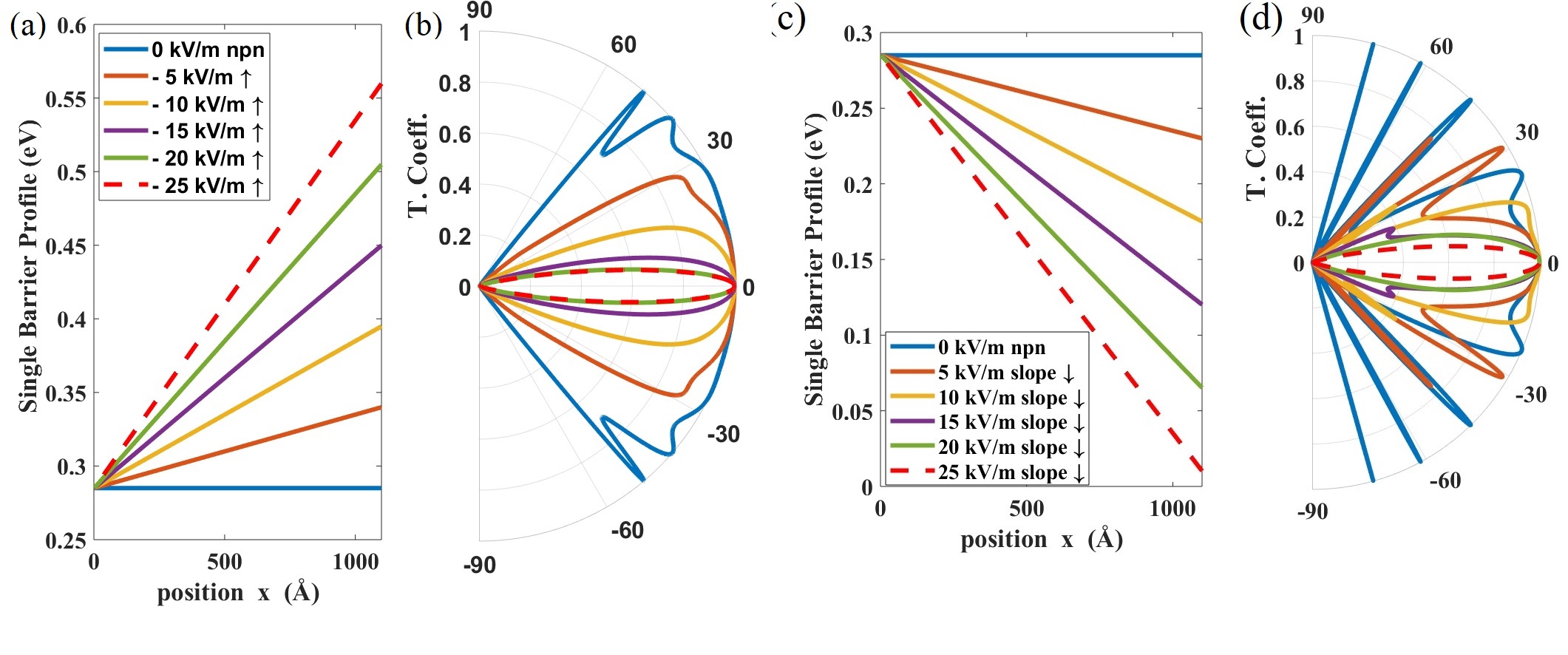}
\caption{(Color online) Polar plots for transmission coefficient $T(\varepsilon,\phi_{\bf k}\,\vert\,{\cal E}_{dc})$ as a function of incident angle $\phi_{\bf k}$
for different bias values ${\cal E}_{dc}$, where both results for an enhanced potential barrier ${\cal E}_{dc}<0$ (left) and a reduced potential barrier ${\cal E}_{dc}>0$ (right) 
are shown in this figure for a full comparison. Here, $\varepsilon=400\,$meV (left) and $80\,$meV (right) are chosen.
The other parameters are the same as those in Fig.\,\ref{f2}.}
\label{f3}
\end{figure} 

\begin{figure}
\centering
\includegraphics[width=0.85\textwidth]{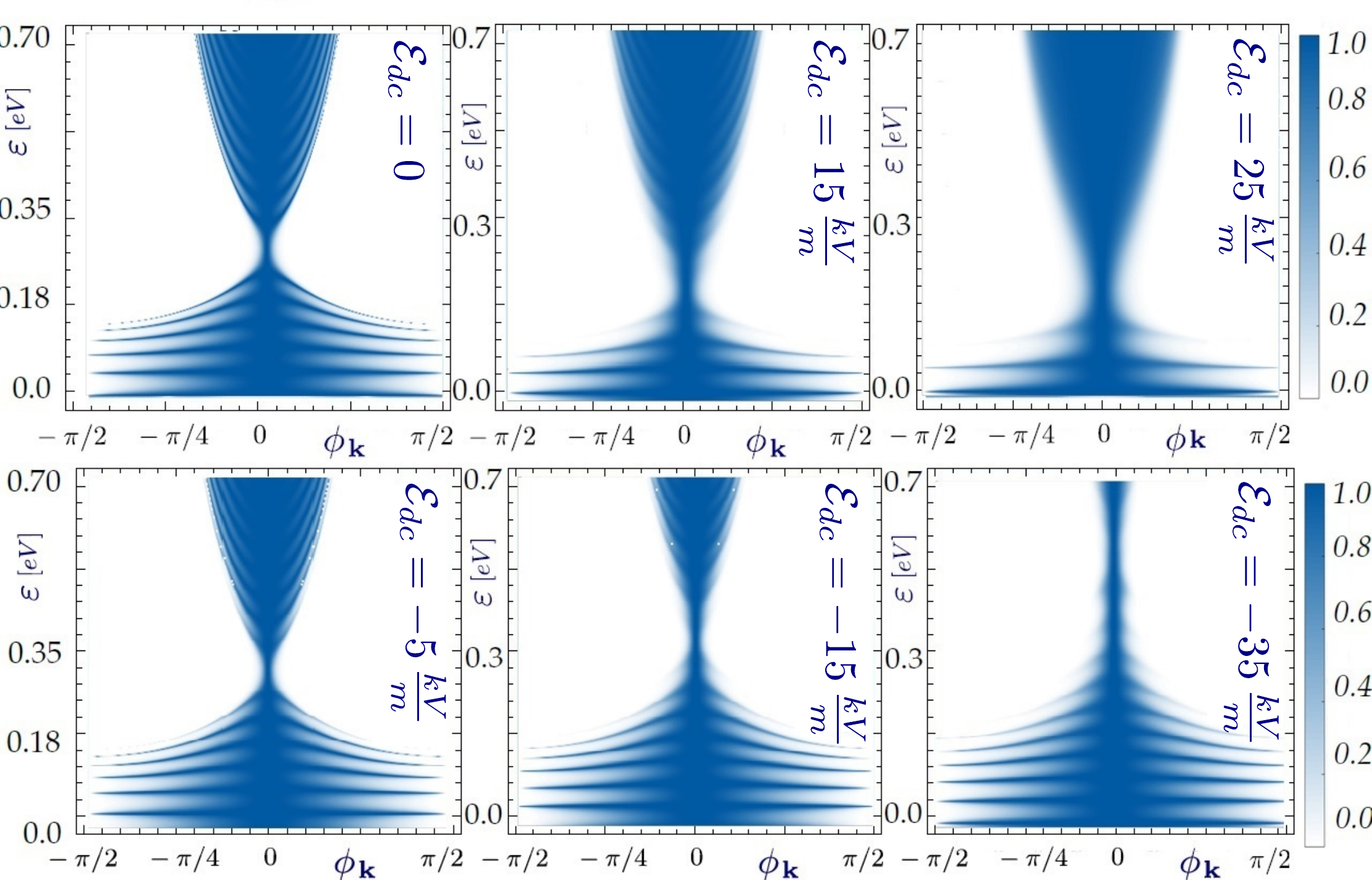}
\caption{(color online) Density plots of $T(\varepsilon,\phi_{\bf k}\,\vert\,{\cal E}_{dc})$ as functions of both $\varepsilon$ and $\phi_{\bf k}$, 
$W_B=110\,$nm, $V_0=285\,$meV and different values of ${\cal E}_{dc}$ are assumed.}
\label{f4}
\end{figure}

\begin{figure}
\centering
\includegraphics[width=0.99\textwidth]{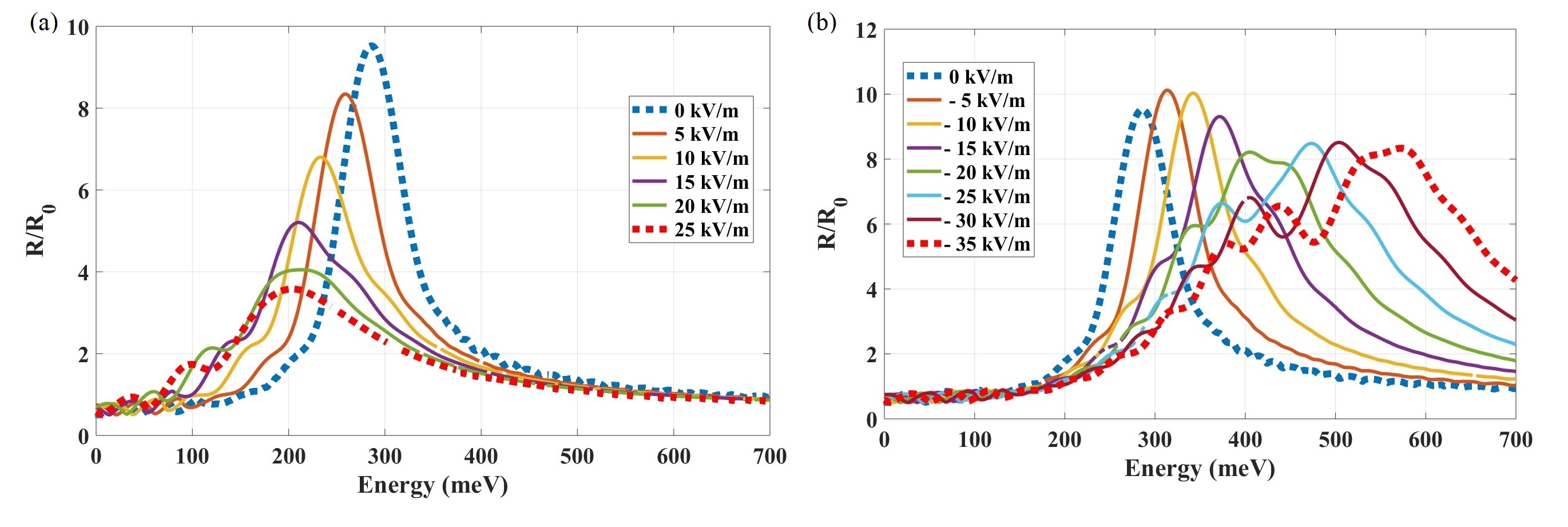}
\caption{(Color online) Ratio of tunneling resistance $R/R_0$ (inverse conductance) 
for ballistic electrons in graphene, calculated from Eq.\,\eqref{cond}, through a biased potential barrier with different values of bias fields. 
Here, $R_0$ is the resistance for normal incidence with $\phi_{\bf k}=0$ and results for both positive (left) and negative (right) biases are presented for comparisons.}
\label{f6}
\end{figure}

The numerical results of $T(\varepsilon,\phi_{\bf k}\,\vert\,{\cal E}_{dc})$
based on finite-difference method are presented in Fig.\,\ref{f3} as a function of incident angle $\phi_{\bf k}$ for various values of bias field ${\cal E}_{dc}$.
Our results indicate that the Klein paradox, i.e., $T(\varepsilon,\phi_{\bf k}\,\vert\,{\cal E}_{dc})=1$ at $\phi_{\bf k}=0$, persists for all considered bias field values, either positive or negative. 
When $|{\cal E}_{dc}|$ is very small, large-angle resonant tunneling only occurs for ${\cal E}_{dc}>0$ or a reduced potential barrier but not for ${\cal E}_{dc}<0$ or an enhanced potential barrier. 
As $|{\cal E}_{dc}|$ becomes large, however, resonant tunneling are squeezed into a narrow angle region around $\phi_{\bf k}=0$ (see Fig.\,\ref{f2} for a comparison). 
Such variations observed in $T(\varepsilon,\phi_{\bf k}\,\vert\,{\cal E}_{dc})$ can be attributed to the modification of a barrier potential profile $V_B(x)$ by a bias field
compared with a square potential barrier $V_B(x)=V_0$.
\medskip

Figure\ \ref{f4} displays density plots of $T(\varepsilon,\phi_{\bf k}\,\vert\,{\cal E}_{dc})$ as functions of both incident energy $\varepsilon$ and incident angle $\phi_{\bf k}$  
with six different values for bias field ${\cal E}_{dc}$. We take the case with ${\cal E}_{dc}=0$ as a starting point, where the Klein paradox and collimation effect exist with 
many sharp resonances (branching and needling features) observed. As ${\cal E}_{dc}$ increases from zero to $250\,$V/cm, these resonant branching and needling features are greatly
obscured although the Klein paradox persists. On the other hand, as negative ${\cal E}_{dc}$ increases from zero to $-350\,$V/cm, both branching and needling regions expand significantly 
to higher incident energy range of electrons.
\medskip

The calculated tunneling coefficient $T(\varepsilon,\phi_{\bf k}\,\vert\,{\cal E}_{dc})$ can be put into Eq.\,\eqref{cond} to find tunneling conductivity or resistivity (its inverse) 
of ballistic electrons through a biased potential barrier in graphene. Here, the conductivity strongly depends on the bias field ${\cal E}_{dc}$ due to nonlinear nature of tunneling transport.
For ballistic Dirac electrons in the absence of a potential barrier, their conductivity should be integer multiple of $2e^2/h$, as indicated by Eq.\,\eqref{cond}. In the presence of scattering by 
impurities or phonons, the occurring resistive force can give rise to a bias-dependent conductivity which is accompanied by a joule heating of electrons. Here, however, 
a bias-dependent conductivity is induced by a tunneling barrier which elastically and coherently reflects incoming electrons, leading to a destructive interference. Such a behavior can be attributed to 
a strong bias modulation ($\alpha$ dependence) of tunneling coefficient $T(\varepsilon,\phi_{\bf k}\,\vert\,{\cal E}_{dc})$ in Eq.\,\eqref{cond}. 
\medskip 

In Fig.\,\ref{f6}, we present the calculated resistance ratios as functions of incident electron energy for different positive and negative biased potential barriers. 
We find from this figure that the resistance peak height decreases with increasing positive ${\cal E}_{dc}$ and the peak position shifts down to lower incident energy $\varepsilon$ at the same time.
For increasing negative ${\cal E}_{dc}$, on the other hand, the peak position shifts upward with $\varepsilon$ but the peak height remains nearly unchanged.
Furthermore, the resistance peak is broadened with increasing $|{\cal E}_{dc}|$, and the broadening effect becomes stronger for negative ${\cal E}_{dc}$ values.
\medskip

\begin{figure}
\centering
\includegraphics[width=0.6\textwidth]{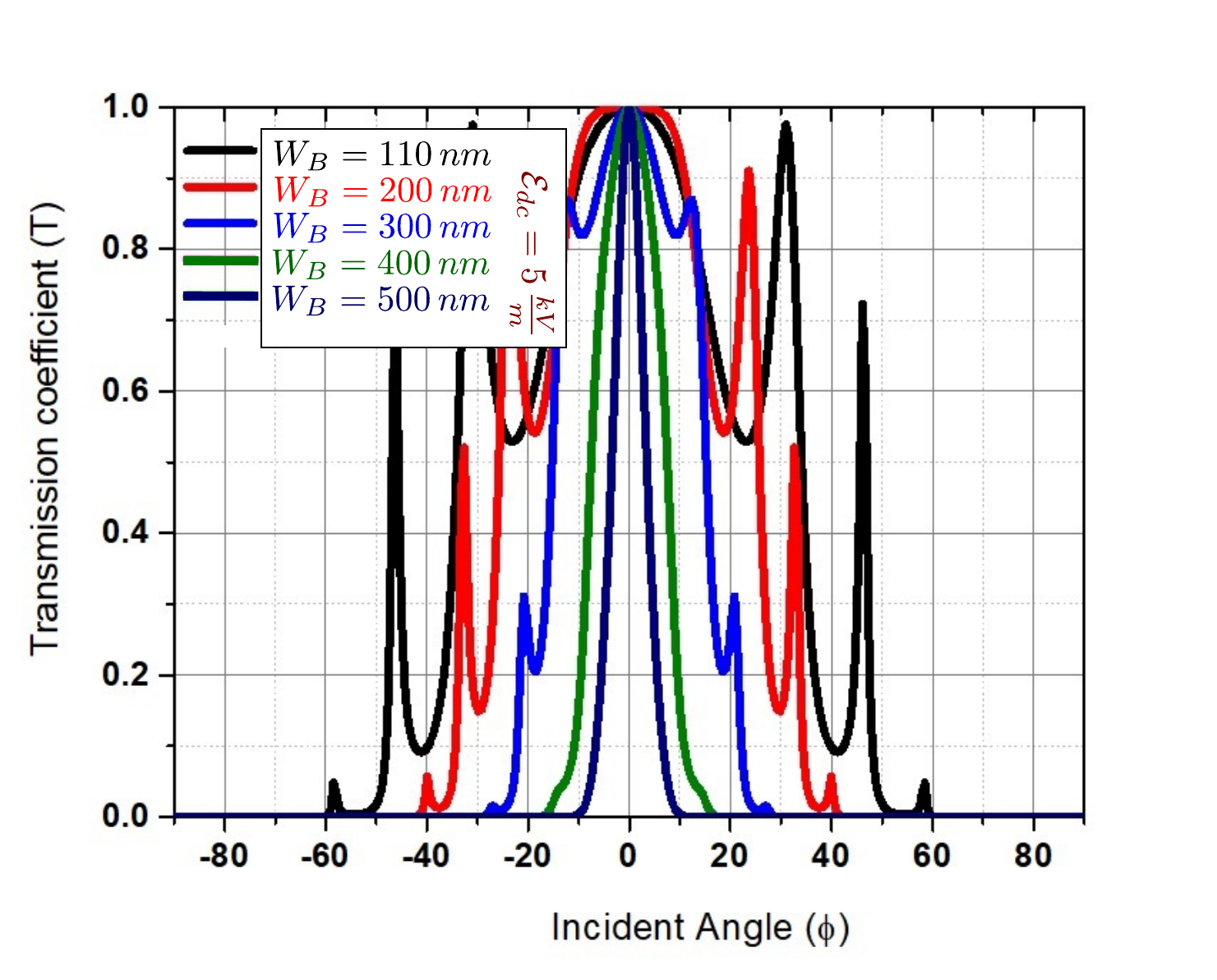}
\caption{(Color online) Transmission coefficient $T(\varepsilon,\phi_{\bf k}\,\vert\,V_0,\alpha)$ at 
${\cal E}_{dc}=50\,$V/cm as a function of incident angle $\phi_{\bf k}$ for various barrier widths $W_B$.}
\label{f5}
\end{figure}

As shown in Fig.\,\ref{f5}, the transmission coefficient $T(\varepsilon,\phi_{\bf k}\,\vert\,{\cal E}_{dc})$ at ${\cal E}_{dc}=50\,$V/cm is suppressed only for large incident angles $|\phi_{\bf k}|$ 
with increasing barrier width $W_B$ due to enlarged switching from a semi-classical regime to a quantum-tunneling regime inside barrier region, as well as due to
interference effect in reflections from both barrier edges. Meanwhile, $T(\varepsilon,\phi_{\bf k}\,\vert\,{\cal E}_{dc})$ for 
$|\phi_{\bf k}|$ around zero remains unchanged, leading to enhanced collimation of Dirac-electron tunneling.

\section{WKB Approximation for Wave Function and Transmission}
\label{s3}

In Section\ \ref{s2}, we have demonstrated a finite-difference approach for calculating 
Dirac-electron tunneling through an arbitrary potential barrier. 
In order to gain physics behind nonlinear transport of 
ballistic Dirac electrons tunneling, we introduce WKB approximation 
so as to analyze the dynamics of tunneling Dirac electrons in an explicit form beyond the 
numerical solution.
\medskip
  
For this purpose, 
let us consider a tilted potential barrier, as shown in Fig.\,\ref{AddFig}, with the potential
$V_B(x)=V_0+\alpha x$, while $V_B(x)=0$ outside of the barrier region.
For this case, the effective $x-$dependent wave vector $k(x)$ can be written as

\begin{equation}
k(x)=\frac{\varepsilon-V_B(x)}{\hbar v_F}=\frac{\varepsilon-V_0}{\hbar v_F}-ax\ ,
\label{barrier}
\end{equation}
where $\varepsilon$ is the energy of an incoming particle, which is conserved for an elastic scattering with the barrier region, and
$a=\alpha/(\hbar v_F)$. For a square potential barrier, as considered in Ref.\,[\onlinecite{neto}], we simply set $a=0$ and will
use it to build up our perturbation theory below.  
\medskip

We first introduce a unitary transformation for a gapless Dirac Hamiltonian, i.e., a $\pi/2$-rotation around the $x-$axis, as employed
in Ref.\,[\onlinecite{sonin2009effect}], for simplification. This leads to the mixed eigen-function $\Phi(x \, \vert \, k_y)=\left[\phi_+,\,\phi_-\right]^T$, 
where $\phi_\pm=(\phi_B\pm\phi_A)/\sqrt{2}$. If $V_B(x)\equiv V_0$ for a constant potential barrier, 
we find the eigen-function

\begin{equation}
\label{Psi0}
\Psi_0(x,\,y\,\vert\,k_y)=
\left[
\begin{array}{c}
\phi_+\\
\\
\phi_-
\end{array}\right]\,\tet{e}^{ik_yy}=
\frac{1}{2}\left[
\begin{array}{c}
\tet{e}^{i \theta_{\bf k}}+1\\
\\
\tet{e}^{i \theta_{\bf k}}-1
\end{array}
\right]\,\tet{e}^{ik_x^{(0)}x+ik_yy}\ ,  
\end{equation}
where $\theta_{\bf k}=\tan^{-1}(k_y/k_x^{(0)})$ is the in-plane angle in the momentum space,
$k_x^{(0)}=\sqrt{[(1/\hbar v_F)(\varepsilon-V_0)]^2-k_y^2}\,$, and the wave-function amplitude is independent of $x$ and $y$. 
\medskip

In the most general case, the wave-function amplitudes $\psi_\pm(x,\,y)$ satisfy the following equations\,\cite{sonin2009effect}
\begin{equation}
\label{Sonin6Main}
\mp i\,\frac{\pr \psi_\pm}{\pr x} \mp \frac{\pr \psi_\mp}{\pr y} =  k(x) \, \psi_\pm\ .
\end{equation}
Throughout our derivation, the translational symmetry in the $y$ direction is always kept since our potential $V_B(x)$ varies only
along the $x-$axis. Therefore, we can simply write down $\psi_{\pm}(x,\,y)=\tet{exp}(ik_yy)\,\phi_{\pm}(x)$.
This simplify Eq.\,\eqref{Sonin6Main} into 

\begin{equation}
\label{simphi}
\mp i\,\partial_x\phi_{\pm}(x)\mp i k_y \,\phi_{\mp}(x) = k(x) \, \phi_{\pm}(x) \ , 
\end{equation}
where $\partial_x\phi(x)\equiv d\phi(x)/dx$. As a special case, one can easily verify that the
solution $\Psi_0(x,\,y\,\vert\,k_y)$ in Eq.\,\eqref{Psi0} satisfies the above equation as $V_B(x)=V_0$ or $a=0$ in Eq.\,\eqref{barrier} is taken for $k(x)$. 

\subsection{WKB Semi-Classical Approach}

The general form of semi-classical WKB expansion for a tunneling-electron wave function
$\Psi(x,\,y\,\vert\,k_y)$ can be expressed as\,\cite{zalip} 

\begin{equation}
\Psi(x,\,y\,\vert\,k_y)= \tet{e}^{(i/\hbar)\,S_\Delta(x)}\,\sum\limits_{s=0}^{\infty}(-i\hbar)^s\, 
\Psi_s(x,\,y\,\vert\,k_y)\ , 
\label{expan}
\end{equation}
where $\displaystyle{S_\Delta(x)=\hbar\int\limits_x d\xi\,k_x(\xi)}$ represents an action.
Here, we will only consider the leading $s=0$ term in Eq.\,\eqref{expan} and obtain
 
\begin{equation}
\Psi_\pm(x,\,y\,\vert\,k_y)=\frac{1}{2} \,\mbb{C}_\Delta (x \,\vert \, k_y) \,
\left(\tet{e}^{i\Theta_\Delta (x\,\vert\,k_y)}\pm 1\right)\,
\tet{exp}\left[\frac{i}{\hbar}\,S_\Delta(x)\right]\,\tet{e}^{ik_yy}\ ,
\label{expan-2}
\end{equation}
where $k_x(\xi)=(1/\hbar v_F)\sqrt{[\varepsilon-V_B(\xi)]^2-(\hbar v_F k_\Delta)^2\,}$ and
$k_\Delta=(1/\hbar v_F)\sqrt{(\hbar v_F k_y)^2+\Delta_G^2\,}$ is independent of $\xi$.
Furthermore, we have also introduced the following two dimensionless quantities in Eq.\,\eqref{expan-2}

\begin{eqnarray}
&& \mbb{C}_\Delta(x\,\vert\,k_y)= 
\frac{1}{k_x(x)}\,\left\{
k_-(x)+i\,\frac{\Delta_G[\varepsilon-\Delta_G-V_B(x)]}{(\hbar v_F)^2\,k_y} 
\right\}\ ,\\
\nonumber 
&& \Theta_\Delta (x\,\vert\,k_y)=\tan^{-1}\left[
\frac{\hbar v_F\,k_+(x)}{\varepsilon-\Delta_G-V_B(x)}
\right]\ ,
\end{eqnarray}
where $k_{\pm}(x)=k_x(x)\pm ik_y$.
It is straightforward to verify that the above solution becomes equivalent to that of gapless graphene as $\Delta_G=0$, given by\,\cite{sonin2009effect}

\begin{equation}
\Psi_\pm(x,\,y\,\vert\,k_y)=\frac{k_+(x)\pm k(x)}{2\sqrt{\vert k(x)\vert\, k_x(x)}}\, \tet{exp} \left[i\int\limits^x d\xi\,k_x(\xi)\right]\,\tet{e}^{i k_y y}\ .
\end{equation}
where $k(x)$ has already been given by Eq.\,\eqref{barrier}.
\medskip

As an electron moves uphill with increasing potential, the sum of its potential and kinetic energies remains as a constant. Therefore, the kinetic energy of the electron decreases on its way. 
For this case, we need define a {\it turning point of a semi-classical trajectory}, at which $k_x(x)=0$ but the total kinetic energy is still positive due to $k_y\neq 0$. We first find the turning point $x_0=(\varepsilon-V_0)/\alpha$ from $k(x)=0$ in Eq.\,\eqref{barrier}, where a Dirac electron turns into a Dirac hole. 
Moreover, the range corresponding to $\vert x-x_0\vert<\xi_c$ becomes a classically forbidden
region in which $k_x(x)$ become imaginary, where $\xi_c=\hbar v_Fk_y/\alpha$ for $\Delta_G=0$. If this forbidden region lies entirely within the biased potential barrier region, 
the transmission coefficient $T (\alpha\,\vert\,k_y)$ is found to be\,\cite{zalip} 

\begin{equation}
\label{wkb1}
T(\alpha\,\vert\,k_y)\backsim\tet{exp}\left[ -2
\int\limits_{x_0-\xi_c}^{x_0+\xi_c} d\xi\,\sqrt{k_y^2\,}\right]
=\tet{exp}\left(-\frac{4}{a}\,k_y^2\right)\ , 
\end{equation}
which is a clear manifestation of the conservation of the Klein paradox for a biased potential barrier layer. 
\medskip

For the case with $\Delta_G>0$, the result in Eq.\,\eqref{wkb1} could be generalized to 

\begin{eqnarray}
\label{wkb2}
&& T_\Delta (\alpha\,\vert\,k_y)\backsim 
\tet{exp} \left[-2\int\limits_{x_0-\xi_c}^{x_0+\xi_c} d\xi\,\sqrt{k^2_\Delta(\xi)\,}
\right]=\tet{exp}\left\{-\frac{4}{a}\left[k_y^2+(\Delta_G/\hbar v_F)^2\right]\right]
\, , \\
\nonumber 
&& \xi_c=\frac{1}{\alpha}\sqrt{(\hbar v_Fk_y)^2+\Delta_G^2\,}\ .
\end{eqnarray}
Therefore, Klein paradox will not exist for any $\alpha$ and $k_y$ values. In addition, an exact solution for the wave function in this case could also be obtained by 
using the results in Ref.\,[\onlinecite{sonin2009effect}], as demonstrated in Appendix\ \ref{apa}.

\subsection{Perturbative Solution at Low Bias}

We would like to emphasize that 
all the results obtained in previous subsection suffer a limitation, i.e., they are valid only if the electron-to-hole switching occurs inside the barrier region. However, this becomes invalid if either the slope $\alpha$ of a potential profile or the barrier width $W_B$ becomes very small. 
\medskip

To seek for a perturbative solution within a barrier layer, we first assume a very small slope $\alpha$ to ensure $a=\alpha/\hbar v_F\ll k^2(x)$. We further assume $\varepsilon>V_B(x)$ so that particle-hole switching will not occur. As a result, the wave function takes the form $\psi_{A,B}(x,y)=\phi_{A,B}(x)\,\tet{exp}(ik_yy)$ and Eq.\,\eqref{simphi} can be applied to find solution for $\phi_{A,B}(x)$. In this case, however, a $\pi/2$-rotation for $\phi_{\pm}(x)$ is not needed. 
\medskip

For $V_B(x)=V_0+\alpha x$, the electron momentum is $k(x)=k_0-ax$, where $k_0=(\varepsilon-V_0)/\hbar v_F$ and $a=\alpha/\hbar v_F$. From this, we find $\partial_xk(x)=-a$, which becomes a small parameter in expansion. Based on these assumptions, we acquire a second-order differential equation with respect to the first wave-function component $\phi_{A}(x)$, yielding

\begin{equation}
\label{mainphi}
\partial^2_x\phi_A(x)+\frac{a}{k_0-ax}\,\partial_x\phi_A(x)+\left[
(k_0-ax)^2-\frac{ak_y}{k_0-ax}-k_y^2\right]\phi_A(x)=0\ .
\end{equation}
Considering the fact that $|a|\ll 1$, we approximate the above equation as

\begin{equation}
\partial^2_x\phi_A(x)+\left(\frac{a}{k_0}+\frac{a^2x}{k_0^2}\right)\partial_x\phi_A(x)+\left[k_0^2-k_y^2-a\left(\frac{k_y}{k_0}+2k_0x\right)+a^2 \left(x^2-\frac{xk_y}{k_0^2}\right)\right]\phi_A(x)=0\ . 
\label{mainphi-1}
\end{equation}
\medskip

Now, we look for a perturbative solution of Eq.\,\eqref{mainphi-1} in the form of $\phi_A(x)=\phi^{(0)}_A(x)+a\,\phi^{(1)}_A(x)+a^2\,\phi^{(2)}_A(x)+\,\cdots\,$, and include only the terms up to the first non-vanishing linear correction to $\phi_A(x)$. Therefore, we get the $0$th and $1$st order equations, respectively, 

\begin{eqnarray}
\label{pert1}
&& a^0: \hskip0.2in \partial^2_x{\phi}^{(0)}_A(x)+\left(k_0^2 -k_y^2\right)\phi^{(0)}_A(x)=0\ ,\\
\label{pert11}
&& a^1: \hskip0.2in \partial_x^2{\phi}^{(1)}_A(x)+\left(k_0^2-k_y^2\right)\phi^{(1)}_A(x)+\frac{1}{k_0}\, \partial_x{\phi}^{(0)}_A(x)-\left(\frac{k_y}{k_0}+ 2k_0x\right)\phi^{(0)}_A(x)=0\ .
\end{eqnarray}
Moreover, making use of the relation in Eq.\,\eqref{simphi} for two components of the wave function, i.e.,

\begin{equation}
\label{phiB}
\phi_B(x) = \frac{\partial_x{\phi}_A(x)-k_y\,\phi_A(x)}{i (k_0-ax)}\equiv\phi^{(0)}_B(x)+a\,\phi^{(1)}_B(x)\ , 
\end{equation}
we find

\begin{eqnarray}
\label{phiB0}
&& \phi^{(0)}_B(x)=\frac{- i}{k_0}\left[\partial_x{\phi}^{(0)}_A(x)-k_y\,\phi^{(0)}_A(x)\right]\ , \\
\label{phiB1}
&& \phi^{(1)}_B(x)=\frac{-i}{k_0^2}\left\{k_0
\left[\partial_x{\phi}^{(1)}_A(x)-k_y\,\phi^{(1)}_A(x)\right]+x\left[\partial_x{\phi}^{(0)}_A(x)-k_y\phi^{(0)}_A(x)\right]\right\}\ . 
\end{eqnarray}
\medskip

For the $0$th order solution, we are dealing with the bias-free case having $a = 0$ or a square potential barrier $V_B(x)=V_0$. From Eq.\,\eqref{pert1} we easily find its solution

\begin{equation}
\label{phiAR}
\phi^{(0)}_A(x)=c_1^{(0)}\tet{e}^{i k_x x}+c_2^{(0)}\tet{e}^{-ik_xx} \hskip0.2in  \mbox{with $k_x= \sqrt{k_0^2-k_y^2\,}$}\ ,  
\end{equation}
which 
is a superposition of the forward and backward plane waves.\,\cite{neto} In this case, from Eq.\,\eqref{phiB0} the corresponding solution for the second component of the wave function is given by 

\begin{eqnarray}
\label{phiBR}
\phi^{(0)}_B(x) &&=c_1^{(0)}\left(\frac{k_x+ik_y}{\gamma\,k_0}\right)\, \tet{e}^{ik_xx}+c_2^{(0)}\left(\frac{-k_x+ik_y}{\gamma\,k_0}\right)\,\tet{e}^{-ik_xx} \\
\nonumber 
&&\equiv\gamma\left(c_1^{(0)}\,\tet{e}^{i\theta_{\bf k}}\,\tet{e}^{ik_xx}-c_2^{(0)}\,\tet{e}^{-i\theta_{\bf k}}\,\tet{e}^{-ik_xx}\right)\ , 
\end{eqnarray}
where $\gamma=\text{sign}\left(\varepsilon-V_0\right)=\pm 1$ is the electron-hole index within the barrier region and $\theta_{\bf k}=\tan^{-1}\left(k_y/k_x\right)$ for Dirac electrons inside 
the barrier region. 
Assuming $\varepsilon>V_0$, we always have $\gamma>0$ and no electron-hole switching will occur. Two constants $c_1^{(0)}$ and $c_2^{(0)}$ in Eq.\,\eqref{phiBR} can be 
determined by boundary conditions at both sides of a barrier layer.
\medskip 

The incoming wave function can be written as\,\cite{neto, oura, ourt} 

\begin{equation}
\label{iw}
\Phi_i(x)=\frac{1}{\sqrt{2}}\left[
\begin{array}{c}
\tet{e}^{i\phi_{\bf k}/2}\\
\\
\tet{e}^{-i\phi_{\bf k}/2} 
\end{array} 
\right]\tet{e}^{ik^{(0)}_xx}\ , 
\end{equation}
where $V_B(x<0)=0$, $k^{(0)}_x=\sqrt{(\varepsilon/\hbar v_F)^2-k_y^2}$ and $\phi_{\bf k}=\tan^{-1}\left(k_y/k_x^{(0)}\right)$ is the incident angle of Dirac electrons. 
Here, the transversal electron wave vector $k_y$ remains to be a constant during the whole tunneling process.
\medskip

We first notice that $c_1^{(0)}$ and $c_2^{(0)}$ in Eqs.\,\eqref{phiAR} and \eqref{phiBR} are not normalized, and they can be determined by the first boundary condition at $x=0$, 
giving rise to 

\begin{eqnarray}
 \nonumber 
 c_1^{(0)}&=&\frac{
\left[\gamma+\tet{e}^{i\left(\phi_{\bf k}+\theta_{\bf k}\right)}\right]\,\left(1+\tet{e}^{2i\phi_{\bf k}}\right)}
{\mbb{D}(k_x, \phi_{\bf k}\,\vert\,\theta_{\bf k})}\ , \\ 
\nonumber 
c_2^{(0)}&=&\frac{\tet{e}^{i\left(2k_xW_B+\theta_{\bf k}\right)}\,\left(1+\tet{e}^{2i\phi_{\bf k}}\right)\,
\left(\gamma\,\tet{e}^{i\theta_{\bf k}}-\tet{e}^{i\phi_{\bf k}}\right)}
{\mbb{D}(k_x, \phi_{\bf k}\,\vert\,\theta_{\bf k})}\ , \\
\nonumber 
\mbb{D}(k_x, \phi_{\bf k}\,\vert\,\theta_{\bf k})&=& 
\gamma+\tet{e}^{i(\phi_{\bf k}+\theta_{\bf k})}\left[2+\gamma\,\tet{e}^{i(\phi_{\bf k}+\theta_{\bf k})}\right]\\
\label{c12}
&+&2\,\tet{e}^{i(2k_xW_B+\phi_{\bf k}+\theta_{\bf k})}
\left[\gamma\cos\left(\theta_{\bf k}-\phi_{\bf k}\right)-1\right]\ .
\end{eqnarray}
Here, $c_1^{(0)}$ and $c_2^{(0)}$ in Eq.\,\eqref{phiAR} play the role of transmission and reflection amplitudes within the barrier region. 
Using the result in Eq.\,\eqref{c12} and the second boundary condition at $x=W_B$ as well, we can further calculate the transmission coefficient $t^{(0)}$ as 

\begin{equation}
\label{t0}
t^{(0)}=\frac{\gamma\,\tet{e}^{-ik_x W_B}\,\cos\theta_{\bf k}\,\cos\phi_{\bf k}}
{\gamma\,\cos (k_x W_B)\,\cos\theta_{\bf k}\,\cos\phi_{\bf k}+i\sin(k_x W_B)\,\left(
\gamma\,\sin\theta_{\bf k}\,\sin \phi_{\bf k}-1\right)}\ ,
\end{equation}
which is identical to the corresponding results for a square potential barrier in Ref.\,[\onlinecite{neto}], as expected. 
\medskip 

In a similar way, we can find the $1$st order solution from Eq.\,\eqref{pert11} for $\phi^{(1)}_A(x)$, yielding 

\begin{eqnarray}
\nonumber 
\phi^{(1)}_A(x)&=&c_1^{(1)}\tet{e}^{ik_xx}+c_2^{(1)}\tet{e}^{-ik_xx}+\mc{F}(x\,\vert\,k_x,\theta_{\bf k})\ \ \ \ \ \ \ \ \mbox{with}\\ 
\nonumber 
\mc{F}(x\,\vert\,k_x,\theta_{\bf k})&=&\frac{\tet{e}^{-ik_xx}}{4k_x^3}\left\{
c_2^{(0)}\Big(-k_x\tet{e}^{-\theta_{\bf k}}\left(2k_xx-i\right)+k_0\left[2k_xx\left(1+ik_xx\right)-i\right]\Big)\right.  \\
\label{phiA1} 
&+&i\,c_1^{(0)}\tet{e}^{2ik_xx}\left.\Big(
-ik_x\tet{e}^{+\theta_{\bf k}}\left(2k_xx+i\right)+k_{0}\left[2k_xx\left(k_xx+i\right)-1\right]\Big)\,\right\}\ .
\end{eqnarray}
Here, the two new undetermined constants $c_1^{(1)}$ and $c_2^{(1)}$ are completely different from the zero-order constants 
$c_1^{(0)}$ and $c_2^{(0)}$ in Eq.\,\eqref{c12}, and they represent the first-order corrections to transmission and reflection amplitudes inside the barrier region.   
By using these calculated first wave-function components $\phi^{(0)}_A(x)$ and $\phi^{(1)}_A(x)$ in Eqs.\,\eqref{phiAR} and \eqref{phiA1}, 
it is straightforward to find the second wave-function component $\phi^{(1)}_B(x)$
from Eq.\,\eqref{phiB1} although its explicit expression becomes a bit tedious to write out. 
\medskip 

Now, we are able to determine the coefficients $c_1^{(1)}$ and $c_2^{(1)}$ in Eq.\,\eqref{phiA1} and the correction to the transmission
coefficient $t^{(0)}$ in Eq.\,\eqref{t0}. For this, we would like to write the transmission and reflection coefficients as 
$t=t^{(0)}+a\,t^{(1)}$ and $r=r^{(0)}+a\,r^{(1)}$, corresponding to the wave functions in Eqs.\,\eqref{phiB1} and \eqref{phiA1}.  
Using the two boundary conditions at $x=0$ and $x=W_B$, we arrive at two equations for $r^{(1)}$ and $t^{(1)}$, given by

\begin{eqnarray}
\nonumber
r^{(1)}\left[
\begin{array}{c}
1 \\
\\
-\tet{e}^{-i\phi_{\bf k}}
\end{array} 
\right]&=&\left[
\begin{array}{c}
\phi^{(1)}_A(x=0\,\vert\,k_x,\theta_{\bf k}) \\
\\
\phi^{(1)}_B(x=0\,\vert\,k_x,\theta_{\bf k})
\end{array} 
\right]\ ,\ \ \ \ \ \ \mbox{and} \\ 
\label{finalt} 
t^{(1)}
\left[ 
\begin{array}{c}
1 \\
\\
\tet{e}^{i\phi_{\bf k}}
\end{array} 
\right]\tet{e}^{ik_x^{(0)}W_B}
&=&\left[
\begin{array}{c}
\phi^{(1)}_A(x=W_B\,\vert\,k_x,\theta_{\bf k}) \\
\\
\phi^{(1)}_B(x=W_B\,\vert\,k_x,\theta_{\bf k})
\end{array} 
\right]\tet{e}^{ik_x W_B}\ . 
\end{eqnarray}
Finally, the transmission amplitude $T(\varepsilon,\phi_{\bf k}\,\vert\,V_0,\alpha)$ can be simply found from
$T(\varepsilon,\phi_{\bf k}\,\vert\,V_0,\alpha)=\vert t^{(0)}+at^{(1)}\vert^2$, where 
$t^{(0)}$ and $t^{(1)}$ are given by Eqs.\,\eqref{t0} and \eqref{finalt}, respectively. 
\medskip

\begin{figure}
\centering
\includegraphics[width=0.6\textwidth]{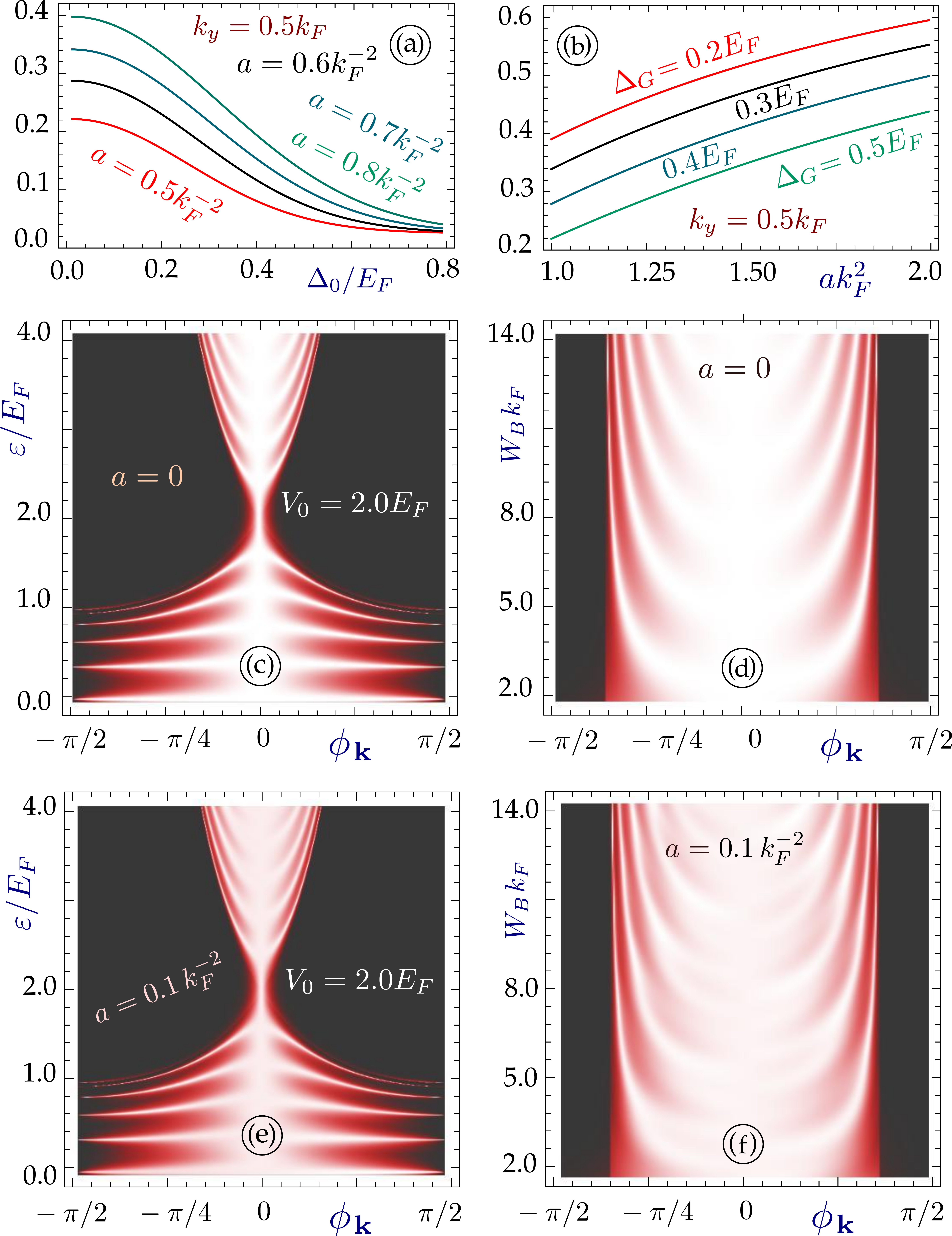}
\caption{(Color online) Transmission amplitude $T(\varepsilon,\,\phi_{\bf k})$ in graphene for fixed $V_0/E_F=2$ and various potential biases specified by different $a$ values. 
Panels $(a)$ and $(b)$ display $T(\varepsilon,\,\phi_{\bf k})$ from Eq.\,\eqref{wkb2} for gapped graphene as functions of $\Delta_G$ and $ak_F^2$, respectively, with fixed $k_y=0.5\,k_F$.
Panels $(c)$-$(e)$ present density plots for $T(\varepsilon,\,\phi_{\bf k})$ from Eqs.\,\eqref{t0} and \eqref{finalt} for gapless graphene $\Delta_G=0$
as functions of $\varepsilon/E_F$ and $\phi_{\bf k}$ in $(c)$, $(e)$ and 
functions of $W_Bk_F$ and $\phi_{\bf k}$ in $(d)$, $(f)$ with $\varepsilon/E_F=1$ for $a=0$ (middle row) and $a=0.1\,k_F^{-2}$ (bottom row).
Here, $E_F=\hbar v_Fk_F=6.28\,meV$ is taken for the energy unit and $k_F$ is the unit for wave vector.}
\label{FIG:AddFig2}
\end{figure}

The numerical results from Eqs.\,\eqref{wkb1} and \eqref{wkb2} for a large electric bias are presented in panels $(a)$ and $(b)$ of Fig.\,\ref{FIG:AddFig2}.
Here, a large graphene gap $\Delta_G$ significantly suppresses $T(\varepsilon,\phi_{\bf k}\,\vert\,V_0,\alpha)$ for all values of $a$, as shown in Fig.\,\ref{FIG:AddFig2}$(a)$, while 
the increase of electric bias $a$ enhances $T(\varepsilon,\phi_{\bf k}\,\vert\,V_0,\alpha)$ for all values of $\Delta_G$, as seen in Fig.\,\ref{FIG:AddFig2}$(b)$.
From Figs.\,\ref{FIG:AddFig2}$(c)$ and \ref{FIG:AddFig2}$(e)$, we find that the full transmission
for a head-on collision remains unchanged even under an electric bias $a\neq 0$. For small $a$ values, the electron-hole transition does not take place
within the barrier region. Instead, a finite $a$ only slightly modifies the resonances of oblique tunneling but not the Klein paradox for the head-on collision.  

\section{Summary and remarks}
\label{s4}

In summary, we have developed a numerical approach for accurately calculating nonlinear tunneling transport of ballistic Dirac electrons through an arbitrary barrier potential in graphene. 
Here, our barrier-potential profile mimics a conventional MOSFET configuration where the square-barrier potential comes from the gate potential, and the 
barrier tilting connects to a source to drain applied bias. In addition, the barrier-potential profile can be tuned by applying a bias, and meanwhile the barrier height can be independently controlled 
by a gate voltage. Our research results can be applied to both sharp and smooth in-plane $p$-$n$ junctions of either bipolar or unipolar devices. 
In the ballistic limit, we propose a mechanism by using barrier profile modulation for Dirac-electron tunneling, allowing an $n$-$p$-$n$ junction to be smoothly converted into
a $p$-$n$ junction with a proper choose of an applied bias.
\medskip

In order to gain insight about the tunneling mechanism of Dirac electrons, we have introduced a perturbation theory for electron transmission through a slanted potential barrier with a small tilting
compared to the inverse barrier width and characteristic electron momenta. 
In addition, we have derived a set of equations, corresponding to different orders of expansion parameter, and obtained analytical solutions of these equations. 
Furthermore, we have demonstrated how the tunneling resonances of a square potential barrier are affected under a finite bias voltage. 
Physically, we have extended a previously developed WKB theory for electron transmission in the opposite limit of a large bias, in which 
electron-to-hole switching occurs within the barrier region. Finally, a finite energy gap in graphene is included and we have
shown that both head-on and skew transmissions will be suppressed exponentially due to existence of an energy gap and a large transverse momentum.  
\medskip 

Another interesting implication is tunable filtering of Dirac electrons for nearly normal incidence,
which might be utilized to design electronic lenses. The uniqueness of our mechanism is that we can specify a range for 
incident electron energies for focusing. Moreover, the electron resistance could be reduced and conductance minima can be shifted in energy just by controlling bias polarity,
barrier height and the strength of bias field. Therefore, our model system can be employed to tune the refractive 
index of such a potential barrier in ballistic-electron optics. All these revealed properties are expected extremely valuable for the development of novel electronic 
and optical graphene-based devices. 

\appendix

\section{Exact Wave Function}
\label{apa}

\begin{figure}
\centering
\includegraphics[width=0.6\textwidth]{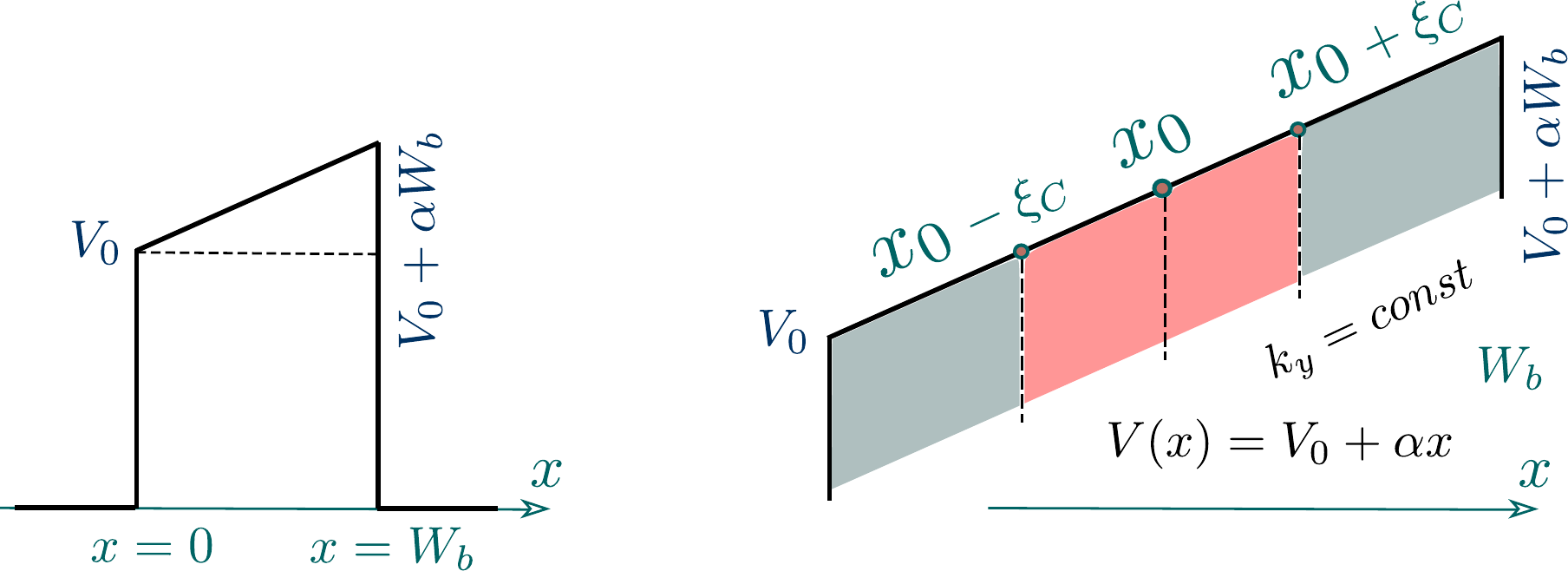}
\caption{(Color online) WKB schematics for the biased potential $V_B(x)=V_0+\alpha x$ in the region of $0\leq x\leq W_B$ and two
classical turning points at $x=x_0\pm\xi_c$.}
\label{AddFig}
\end{figure}

In this Appendix, we seek for an exact solution for the electron/hole wave function in the finite-slope region of 
a barrier, as seen in Fig.\,\ref{AddFig}, with potential $V_B(x)=V_0+\alpha x$. If two boundaries of the barrier region stay 
far away from the electron-to-hole crossing point, i.e., $k(x_0)\approx 0$, the wave function could be 
written as\,\cite{sonin2009effect}

\begin{equation}
\Psi^{(B)} (x\,\vert\,k_y)=\left\{c_1\left[
\begin{array}{c}
\mc{F}(\eta,\zeta) \\
\mc{G}(\eta,\zeta)
\end{array}
\right]+
c_2\left[
\begin{array}{c}
\mc{F}^{\star}(\eta,\zeta) \\
\mc{G}^{\star}(\eta,\zeta)
\end{array}
\right]\, 
\right\}\, \tet{e}^{i k_y y} \ ,
\label{app1}
\end{equation}
where $\eta(x)=(x-x_0)\sqrt{a}$, $\zeta(k_y) = k_y/\sqrt{a}$, the symbol $^\star$ means taking complex conjugation, and 
the two arbitrary constants $c_1$ and $c_2$ will be fixed by the boundary conditions on each side 
of the barrier region. Moreover, two functions $ \mc{F}(\eta,\zeta)$ and $\mc{G}(\eta,\zeta)$ in Eq.\,\eqref{app1} can be expressed by a 
Kummer confluent hypergeometric function $\mc{M}(a, b \, \vert \, z)$ as\,\cite{sonin2009effect}

\begin{eqnarray}
&& \mc{F}(\eta,\zeta)=\tet{exp}\left(-\frac{i}{2}\,\eta^2\right)\, \mc{M}\left(-\frac{i}{4}\,\zeta^2,\,\frac{1}{2}\,\Big|\,i\eta^2\right)\ , \\
\nonumber 
&& \mc{G}(\eta,\zeta)=-\zeta\eta\,\tet{exp}\left(-\frac{i}{2}\,\eta^2\right)\,\mc{M}\left(1-\frac{i}{4}\,\zeta^2,\,\frac{3}{2}\,\Big|\,i\eta^2\right)\ .
\end{eqnarray}
The wave functions outside of the barrier region are easily obtained for the incoming and 
reflected waves, yielding

\begin{equation}
\label{Psi01}
\Psi^{(L)}(x\,\vert\,\mbox{\boldmath$k$})=
\frac{1}{2}\left[
\begin{array}{c}
\tet{e}^{i\phi_{\bf k}}\\
\pm 1
\end{array}
\right]\,\tet{e}^{i k^{(0)}_x x} \, \tet{e}^{i k_y y} +
\frac{r_{\bf k}}{2}\,\left[
\begin{array}{c}
-\tet{e}^{ - i \phi_{\bf k}}\\ 
\pm 1
\end{array}
\right]\,\tet{e}^{-ik^{(0)}_xx}\,\tet{e}^{ik_yy}\ ,  
\end{equation}
where $\phi_{\bf k}=\tan^{-1}(k_y/k_x^{(0)})$.
Similarly, for the transmitted wave we have

\begin{equation}
\label{Psi03}
\Psi^{(R)}(x\,\vert\,\mbox{\boldmath$k$}')=
\frac{t_{{\bf k}'}}{2}\,\left[
\begin{array}{c}
\tet{e}^{i\theta_{{\bf k}'}}\\
\pm 1
\end{array}
\right]\,\tet{e}^{ik_xx}\,\tet{e}^{ik_yy}\ ,
\end{equation}
where $\theta_{{\bf k}'}=\tan^{-1}(k_y/k_x)$.
The transmission coefficient $t_{{\bf k}'}$ and the reflection coefficient $r_{\bf k}$ can be obtained 
by matching the wave functions at two boundaries at $x=0$ and $x=W_B$, i.e., $\Psi^{(L)}(x=0\,\vert\,\mbox{\boldmath$k$})= \Psi^{(B)}(x=0\,\vert\,k_y)$ and $ \Psi^{(B)}(x=W_B\,\vert\,k_y)=\Psi^{(R)}(x=W_B\,\vert\,\mbox{\boldmath$k$}')$. Therefore, we acquire four equations for these two-components wave functions, 
which can be used to determine four unknowns $r_{\bf k}$, $c_1$, $c_2$, and $t_{{\bf k}'}$, and the calculated $t_{{\bf k}'}$ will be further applied for
evaluating the transmission $T_{{\bf k}'}=\vert t_{{\bf k}'}\vert^2$. 
\medskip

Here, we would like to emphasize that although the obtained solution in Eq.\,\eqref{app1} is exact, it holds true only for a very thick potential barrier with $0\ll x_0\ll W_B$. 
Additionally, using this approach we can not address the limiting case with a small slope $a\rightarrow 0$. 
\medskip 

For the boundaries of a very thick potential barrier with a substantial slope $\alpha$, we find vary large absolute value of $\eta(x)=(x-x_0)\sqrt{a}$, and
the wave function is calculated as

\begin{eqnarray}
 && \lim\limits_{\eta\to\infty}\,\Psi^{(B)}(x,\,\vert\,k_y)=\left[ 
 \begin{array}{c}
  0 \\
  1
 \end{array}
 \right]\,\tet{exp}\left[-\frac{i}{2}\eta^2(x)\right]\,\tet{e}^{ik_yy}\ , \\
 \nonumber 
 && \lim\limits_{\eta\to-\infty}\,\Psi^{(B)}(x,\,\vert\,k_y)=\tet{e}^{ik_yy}\,\left\{\tet{exp}\left[-\frac{\pi}{2a}\,k_y^2\right]\, 
 \left[ 
 \begin{array}{c}
  0 \\
  1
 \end{array}
 \right]\,\tet{exp}\left[-\frac{i}{2}\,\eta^2(x)\right]\,+const\,\left[ 
 \begin{array}{c}
  0 \\
  1
 \end{array}
 \right]\, 
 \tet{exp}\left[ 
 \frac{i}{2}\,\eta^2(x)\right]\,\right\}\ , 
\end{eqnarray}
which gives rise to the transmission $T_{{\bf k}'}=\tet{exp}\,\left(-\pi k_y^2/a\right)$. This result is the same as that obtained from the 
a semi-classical theory. 

\bibliography{Farhana}

\end{document}